\pdfoutput=1

\documentclass[aps,prb,reprint,showpacs,superscriptaddress]{revtex4-1}

\usepackage{graphicx}
\usepackage{amsmath}
\usepackage{amssymb}
\usepackage{amsfonts}
\usepackage{dcolumn}
\usepackage{dsfont}
\usepackage{latexsym}
\usepackage{rotating}
\usepackage{color}
\usepackage{latexsym}
\usepackage{bbm}
\usepackage{subfigure}
\usepackage{float}
\usepackage{epsfig}
\usepackage{psfrag}
\usepackage{bm}
\usepackage{amsthm}
\usepackage{eucal}
\usepackage{mathrsfs}
\usepackage{url}
\usepackage{braket}

\usepackage{color} 

\usepackage{hyperref}
\hypersetup{
colorlinks=true,final=true,
        linkcolor=blue,
        citecolor=blue,
        filecolor=blue,
        urlcolor=blue,
}

\begin{document}

\title{Multi-band theory of superconductivity at the LaAlO$_3$/SrTiO$_3$ interface}

\author{N. Mohanta}
\email{nmohanta@phy.iitkgp.ernet.in}
\affiliation{Department of Physics, Indian Institute of Technology Kharagpur, W.B. 721302, India}
\author{A. Taraphder}
\affiliation{Department of Physics, Indian Institute of Technology Kharagpur, W.B. 721302, India}
\affiliation{Centre for Theoretical Studies, Indian Institute of Technology Kharagpur, W.B. 721302, India}

\begin{abstract}
We present a multi-band model for superconductivity at the metallic interface between insulating oxides LaAlO$_3$ and SrTiO$_3$ (001). Using a self-consistent Bogoliubov-de Gennes theory, formulated with the realistic bands at the interface, we investigate the spin-singlet and spin-triplet pairings in intra-band and inter-band channels. We find that the Rashba and atomic spin-orbit interactions at the interface induce singlet pairing in the inter-band channel and triplet pairing in both the intra-band and inter-band channels when the pairing amplitude in the singlet intra-band channel is finite. The gate-voltage variation of superconductivity is resolved in different pairing channels,  compared with experimental results and found to match quite well. Interestingly, an enhancement of the superconducting transition temperature by external in-plane magnetic field is found revealing the existence of a hidden superconducting state above the observed one. As the interface is known to possess high level of inhomogeneity, we explore the role of non-magnetic disorder incorporating thermal phase fluctuations by using a Monte-Carlo method. We show that even after the transition to the non-superconducting phase, driven by temperature or magnetic field, the interface possesses localized Cooper pairs whose signature was observed in previous experiments. 
\end{abstract}

\pacs{74.78.-w, 74.20.Rp, 74.40.-n, 61.43.Bn}

\maketitle

\section{Introduction}
The discovery of superconductivity ($T_c\simeq200$ mK) at the interface~\cite{Reyren31082007,GariglioJPCM2009} between perovskite band insulators LaAlO$_3$ (LAO) and SrTiO$_3$ (STO) triggered a plethora of investigations~\cite{Michaeli_PRL2012,mohantaJPCM,NM_VacancyJPCM2014,Mohanta_EPL2014,Caprara_PRB2013,Kelly_PRB2014,Pavlenko_PRB2009,Vanderbilt_PRB2009,Marel_PRB2014,CNayak_PRB2013} in the last few years due to its exotic nature arising primarily from the presence of competing ferromagnetism ($T_{Curie}\simeq200$ K)~\cite{Li_Nature2011,Dikin_PRL2011,Dagan_PRL2014}, spin-orbit interaction (SOI)~\cite{Fischer_NJP2013,Fete_PRB2012} and disorder~\cite{Ariando_PRX2013,Li_PRB2011,Zhong_PRB2010}. Besides, the quasi-two dimensional electron gas (q2DEG) at the interface~\cite{Ohtomo2004} exhibits intriguing novel properties such as metal-insulator transition~\cite{Thiel_Science2006,Huijben_NMat2006,Cen_NMat2008,Rijnders_Nature2008,Eyert_PRB2013} and ferroelectricity~\cite{Tra_AdMa2013}. On top of that, the ability to control these properties by external electric field~\cite{Caviglia2008, CavigliaPRL2010} added an extra-dimension to the nanoelectronics industry~\cite{Mannhart26032010,Bjaalie_NJP2014}.

The q2DEG is formed by an electronic transfer mechanism in which half an electronic charge per unit cell is transferred to the interface to avoid a polar discontinuity~\cite{Nakagawa2006,Satpathy_PRL2008}. The electrons are confined in a few TiO$_2$ layers located within a region of about 10 nm thickness at the interface and occupy the $t_{2g}$ orbitals of Ti ions~\cite{Pavlenko_PRB2012_1,Pentcheva_PRB2006}. Density functional theory (DFT) reveals that the $d_{xy}$ band is situated below the $d_{yz}$, $d_{zx}$ bands by $\sim0.4$ eV due to the confinement at the interface~\cite{Delugas_PRL2011,Hirayama_JPSJ2012}. The spin-degeneracy of the bands is lifted by the inversion symmetry-breaking Rashba SOI and an atomic SOI~\cite{Held_PRB2013,Khalsa_PRB2013}. Magnetotransport measurements infer the presence of two types of carriers with different mobilities and the high-mobility carriers have been predicted to be responsible for superconductivity~\cite{Shalom_PRL2010,Joshua_PNAS2013,Joshua_NComm2012}. Michaeli \textit{et al.}~\cite{Michaeli_PRL2012} suggested that the system hosts the antagonistic ferromagnetic and superconducting orders by favouring a disordered stabilized helical FFLO state induced by strong Rashba SOI. The complex nature of the coexisting phases~\cite{Li_Nature2011,Dikin_PRL2011,Bert_NPhys2011} has naturally led to the predictions of unconventional superconductivity~\cite{Scheurer2015} and different magnetic ground states. The microscopic understanding of the origin of superconductivity remained obscure until the recent convincing evidence of electron-phonon coupling, obtained using tunneling spectroscopy~\cite{Boschker_arXiv2015}. It is, therefore, apparent that phonons play the dominant role in electron pairing in other STO-based superconductors such as doped STO~\cite{Schooley_PRL1964} and X/STO (X = LaTiO$_3$~\cite{Biscaras2010}, GdTiO$_3$~\cite{Stemmer_PRX2012}, FeSe~\cite{Lee2014}) interfaces. 
The superconductivity at the LAO/STO interface is unique in the following aspects: (i) it appears at very low career concentrations ($\sim 10^{-13}$ cm$^{-2}$)~\cite{Caviglia2008}, (ii) the transition temperature ($T_c$) shows BKT-like behaviour~\cite{Caviglia2008} while the pairing-gap or the superfluid density follows BCS prediction: $2\Delta_{0}/(k_{B}T_{g})\simeq3.4$, where $\Delta_{0}$ is the pairing gap at $T=0$, $k_{B}$ is the Boltzmann constant and $T_g$ is the so called 'gap-closing temperature'~\cite{RichterNature2013}, (iii) it coexists with inhomogenous ferromagnetic puddles of large moments ($\sim 0.4\mu_{B}$ per interface unit cell)~\cite{Li_Nature2011}. The coexistence of the competing orders, albeit in phase segregated regions~\cite{Bert_NPhys2011, mohantaJPCM,NM_VacancyJPCM2014}, gives rise to fascinating phenomena such as the enhancement of superconductivity by magnetic field~\cite{Gardner2011} and the magnetic field assisted transient superconductivity~\cite{Aveek_arxiv2014} leading to possible 'hidden order'. However, despite intensive previous studies, complete theoretical understanding of the nature of the multi-band superconductivity is lacking and necessitates a careful and thorough theoretical analysis.

In the following, we use a three-orbital model for superconductivity to develop an understanding of the nature of superconductivity at the interface in the presence of magnetic moments and try to shed light on the questions raised above. We study the spin-singlet and spin-triplet electron pairing in intra-band and inter-band channels. Using a self-consistent Bogoliubov-de Gennes (BdG) method, formulated with the realistic bands at the interface, we explore the mean-field phase-diagrams, the role of spin-orbit interactions and external magnetic field on the electron pairing. It is found that the pairing in the singlet inter-band channel and triplet intra-band and inter-band channels are induced by the Rashba and atomic SOI when the pairing amplitude in the singlet intra-band channel is finite. Taking cue from the experimental data, we incorporate the gate-voltage in our analysis, study the gate-voltage variation of the pairing amplitudes and plot the phase-diagram to compare with the experimental results. We study the behaviour of the pairing amplitudes in presence of an external in-plane magnetic field and find an enhancement of the superconducting transition temperature when the magnetic field is applied along certain directions in the interface plane. The magnetic field enhancement of superconductivity has been observed experimentally~\cite{Gardner2011} and arises because of the interplay between superconductivity and the competing ferromagnetism. It suggests a hidden superconducting phase above the superconducting transition temperature. Since the interface superconductivity is highly inhomogenous in nature and appears at very low career concentration, the thermal phase-fluctuation becomes  significant. We study the phase-transition from superconductor to a non-superconducting state, driven by temperature or perpendicular magnetic field, taking into account the thermal phase-fluctuation using a Monte-Carlo method and observe that there are localized Cooper-pairs in the non-superconducting phase. The presence of these localized Cooper-pairs has been confirmed in previous experiment~\cite{Mehta2012}. 
 
The remainder of this paper is organized as follows. In Sec.~\ref{model}, we discuss about the origin of ferromagnetism and superconductivity and introduce the multi-orbital effective Hamiltonian for the interface q2DEG. In Sec.~\ref{bdg}, we formulate the self-consistent equations for the superconducting order parameters within the multi-orbital BdG framework. In Sec.~\ref{results}, we present and discuss our results obtained within the self-consistent BdG method. In Sec.~\ref{mc_phase}, we present of Monte-Carlo analysis of the thermal phase-fluctuation in superconductivity. The conclusions are briefly summarized in Sec.~\ref{conclusions}.

\section{Effective three-orbital model of superconductivity}
\label{model}
In the following, we elaborate the microscopic mechanisms of ferromagnetism, superconductivity and other ingredients of the interface q2DEG and establish the effective Hamiltonian for the interface electrons.
 
\begin{figure}[!ht]\vspace{0em}
\begin{center}
\epsfig{file=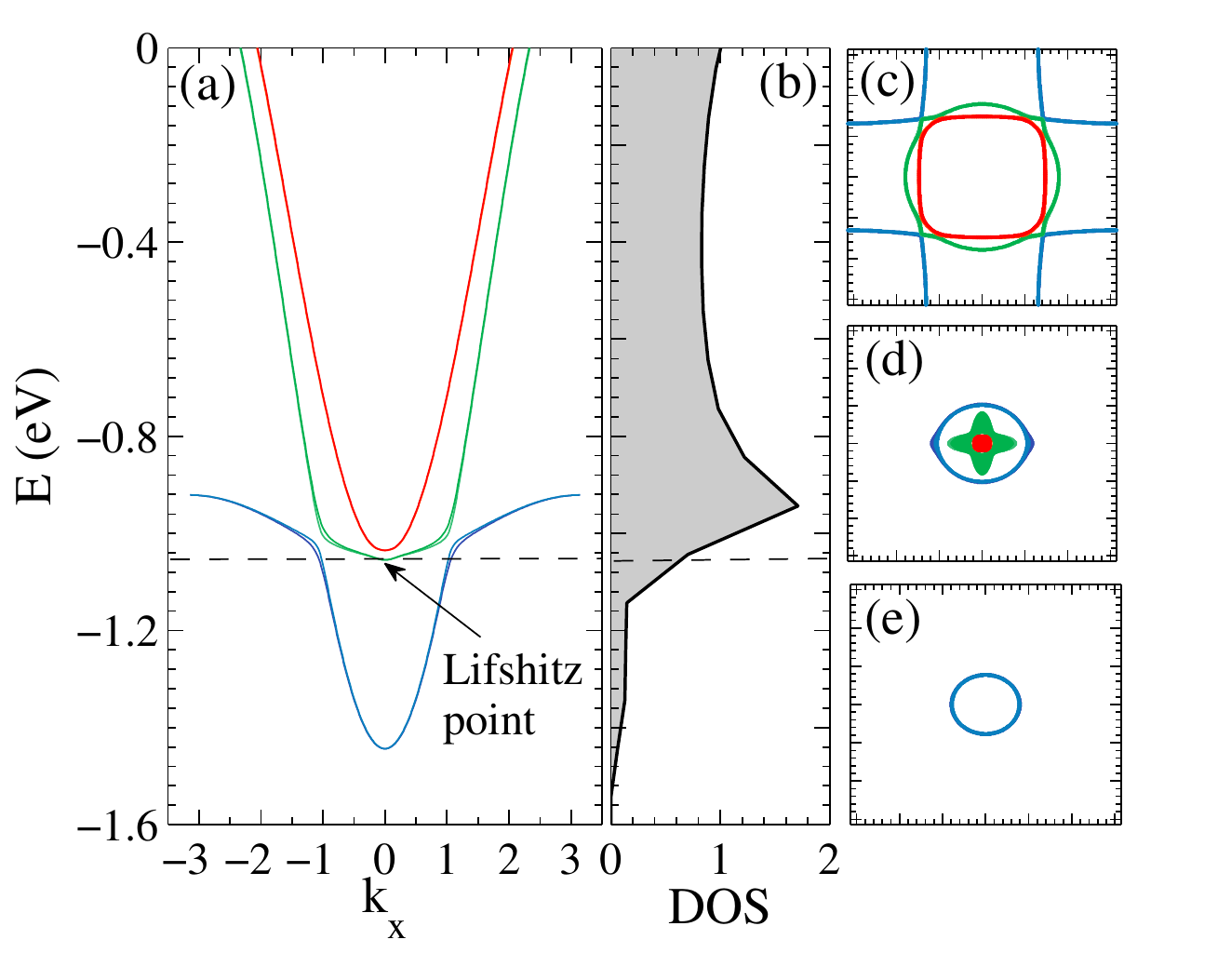,trim=0.05in 0.2in 0.2in 0.1in,clip=true, width=90mm}\vspace{0em}
\caption{(Color online) (a) The band structure of the three $t_{2g}$ orbitals in presence of the Rashba and atomic SOI (spectrum of ${\cal H}_0+{\cal H}_{ASO}+{\cal H}_{RSO}$). There is a mixing of the $d_{xy}$ orbital (blue lines) with the $d_{yz}$ orbital (green lines) and the $d_{zx}$ orbital (red lines) due to the Rashba SOI. The dashed horizontal line, at $\mu\simeq-1.04$ eV, denotes a Lifshitz transition point at which the two upper bands start getting occupied. (b) the total density of states as a function of energy. The Lifshitz transition is reflected by the sharp jump in the density of states near the transition point. (c)-(d)-(e) depicts the change in the Fermi-surface topology at energy below, near and above the Lifshitz transition.}
\label{band_fs}\vspace{0em}
\end{center}
\end{figure}

The electrons coming from the top LaAlO$_3$ layer to neutralize the polarization discontinuity at the interface, predominantly occupy the $d_{xy}$ orbitals of Ti ions in the terminating TiO$_2$ layer and establish quarter-filled $d_{xy}$ states. Because of large onsite Hubbard, and nearest-neighbour Coulomb repulsive interactions at the interface, all the electrons get localized at the interface sites and form a charge-ordered insulating ground state with a weak anti-ferromagnetic super-exchange coupling mediated via the Oxygens~\cite{Michaeli_PRL2012,Banerjee2013}. Additional electrons, supplied by the application of the back-gate voltage or the Oxygen vacancies near the interface, will find the top TiO$_2$ layer as energetically unfavourable and go to the next TiO$_2$ layer to occupy the $t_{2g}$ orbitals of Ti ions. The electrons in the $t_{2g}$ orbitals in the next TiO$_2$ layer participate in conduction and exhibit superconductivity. Spectroscopic experiment~\cite{Joshua_NComm2012} and DFT studies~\cite{Held_PRB2013,Khalsa_PRB2013} show that, due to confinement at the interface, the $d_{xy}$ band is lower in energy at the $\Gamma$-point by $\sim0.4$ eV than the quasi-one dimensional, relatively heavier $d_{yz}$,$d_{zx}$ bands. When the Fermi-level is tuned, the system encounters a Lifshitz transition at an electron concentration $n_c\simeq1.68\times10^{13}$ cm$^{-2}$ where the low-mobility $d_{yz}$,$d_{zx}$ electrons start getting occupied~\cite{Joshua_NComm2012} as depicted in FIG.~\ref{band_fs}.   
An interesting feature of the band-structure is that because of the atomic SOI, the orbital characters of the $d_{xy}$ band and the heavier $d_{yz}$,$d_{zx}$ bands get interchanged near the Lifshitz point and the splitting due to Rashba SOI is significant only near the degeneracy points. The spin-degeneracy in all the bands is lifted by the spin-orbit interactions, the splitting being largest near the band-mixing points.

The wave-functions of the itinerant electrons  in the $t_{2g}$ orbitals in the TiO$_2$ layer below the interface are extended to the terminating TiO$_2$ layer. Therefore, these electrons interact via a ferromagnetic exchange with the localized moments leading to the in-plane ferromagnetic order. The exchange interaction can be described by a Hamiltonian ${\cal H}_{FM}=\sum_i \sum_{\alpha}\int d{\bf r} J_{\alpha}\hat{\bf S}({\bf R}_i)\cdot \hat{\bf s}_{\alpha}({\bf r})\delta({\bf r}-{\bf R}_i)$, where $\hat{\bf S}({\bf R}_i)$ is the spin operator at the local moment sites ${\bf R}_i$, $\hat{\bf s}_{\alpha}({\bf r})$ is the spin-density operator in the itinerant orbital $\alpha$ ($d_{xy},d_{yz},d_{zx}$), $J_{\alpha}$ is the strength of the exchange interaction. Since the itinerant $d_{yz}$,$d_{zx}$ orbitals are orthogonal to the localized $d_{xy}$ orbital, $J_{d_{xy}}\gg J_{d_{yz},d_{zx}}$~\cite{Michaeli_PRL2012}. This is in agreement with the spectroscopic studies which indicate $d_{xy}$ nature of the in-plane ferromagnetism~\cite{Lee_Nature2013}. When treated at the mean-field level, ${\cal H}_{FM}$ will essentially be reduced to 
${\cal H}_{FM}=\sum_{k,\alpha,\sigma,\sigma^{\prime}}( h_{x\alpha} {\sigma_x})_{\sigma,\sigma^{\prime}}c_{k\alpha\sigma}^\dagger c_{k\alpha\sigma^{\prime}}$, where $h_{x\alpha}$ are the Zeeman splitting amplitudes, corresponding to different orbitals of index $\alpha$, along the in-plane direction (taken to be along $\hat{x}$ axis) with $h_{x\alpha}=h_{x1}$ for $d_{xy}$ orbital and $h_{x\alpha}=h_{x2}$ for $d_{yz}$, $d_{zx}$ orbitals ($h_{x1} > h_{x2}$). The attractive interaction, mediated by electron-phonon coupling, for the three itinerant orbitals can be expressed as ${\cal H}_{SC}=-g\sum_{k,k^{\prime},\alpha,\beta}c_{k\alpha\uparrow}^{\dagger}c_{-k\beta\downarrow}^{\dagger}c_{-k^{\prime}\beta\downarrow}c_{k^{\prime}\alpha\uparrow}$, where $g$ is the strength of the pairwise electron-electron interaction. The spin-singlet and spin-triplet pairing amplitudes in the intra-band and inter-band channels can be defined as 
$\Delta_{\alpha\beta}^{s}=-g\braket{c_{k\alpha\uparrow}c_{-k\beta\downarrow}}$ and 
$\Delta_{\alpha\beta\sigma\sigma}^{t}=-g\braket{c_{k\alpha\sigma}c_{-k\beta\sigma}}$ respectively. Neglecting fluctuations beyond mean-field, the pairing term becomes ${\cal H}_{SC}=\sum_{k,\alpha,\beta} (\Delta_{\alpha\beta}^{s}c_{k\alpha\uparrow}^\dagger c_{-k\beta\downarrow}^\dagger + \Delta_{\alpha\beta\uparrow\uparrow}^{t}c_{k\alpha\uparrow}^\dagger c_{-k\beta\uparrow}^\dagger+ \Delta_{\alpha\beta\downarrow\downarrow}^{t}c_{k\alpha\downarrow}^\dagger c_{-k\beta\downarrow}^\dagger + h.c.)$. We have $\Delta_{\alpha\beta\uparrow\uparrow}^{t}=\Delta_{\alpha\beta\downarrow\downarrow}^{t}$ and the orbitals are indexed according to $(a,b,c)=(d_{xy},d_{yz},d_{zx})$.

Another significant feature of the interface q2DEG is the presence of the atomic and Rashba SOI which reorganize the spin and orbital degrees of freedom of the $t_{2g}$ electrons. The atomic SOI, described by the Hamiltonian ${\cal H}_{ASO}=\Delta_{so}\vec{l}\cdot\vec{s}$, appears because of the crystal field splitting of the atomic orbitals. In the $t_{2g}$ orbital basis $(c_{ka\uparrow},c_{kb\uparrow},c_{kc\uparrow},c_{ka\downarrow},c_{kb\downarrow},c_{kc\downarrow})$, ${\cal H}_{ASO}$ can be written as~\cite{Held_PRB2013} 
\begin{align}
&{\cal H}_{ASO}=\frac{\Delta_{so}}{2}\sum_{k}\begin{pmatrix} \begin{array}{cccccc} c_{ka\uparrow}^\dagger & c_{kb\uparrow}^\dagger & c_{kc\uparrow}^\dagger & c_{ka\downarrow}^\dagger & c_{kb\downarrow}^\dagger & c_{kc\downarrow}^\dagger\end{array} \end{pmatrix}\nonumber\\
&\times\begin{pmatrix} \begin{array}{cccccc} 0 & 0 & 0 & 0 & 1 & -i\\0 & 0 & i & -1 & 0 & 0\\0 & -i & 0 & i & 0 & 0\\0 & -1 & -i & 0 & 0 & 0\\1 & 0 & 0 & 0 & 0 & -i\\i & 0 & 0 & 0 & i & 0 \end{array} \end{pmatrix} \begin{pmatrix} \begin{array}{c} c_{ka\uparrow} \\ c_{kb\uparrow} \\ c_{kc\uparrow} \\ c_{ka\downarrow} \\ c_{kb\downarrow} \\ c_{kc\downarrow}\end{array} \end{pmatrix}
\label{H_aso}
\end{align}
where $\Delta_{so}=19.3$ meV is the strength of the atomic SOI. On the other hand, the Rashba SOI, which describes the broken inversion symmetry at the interface, is given by the following Hamiltonian:
\begin{align}
&{\cal H}_{RSO}=\gamma\sum_{k,\sigma}\begin{pmatrix} \begin{array}{ccc} c_{ka\sigma}^\dagger & c_{kb\sigma}^\dagger & c_{kc\sigma}^\dagger\end{array} \end{pmatrix}\nonumber\\
&\times\begin{pmatrix} \begin{array}{ccc} 0 & -2i\sin{k_x} & -2i\sin{k_y}\\2i\sin{k_x} & 0 & 0\\2i\sin{k_y} & 0 & 0\end{array} \end{pmatrix} \begin{pmatrix} \begin{array}{c} c_{ka\sigma} \\ c_{kb\sigma} \\ c_{kc\sigma}\end{array} \end{pmatrix}
\label{H_rso}
\end{align}
where $\gamma=20$ meV is the strength of the Rashba SOI. It is interesting to note that the Rashba SOI, described by ${\cal H}_{RSO}$, is very different from what is usually observed in the 2DEG at semiconducting hetero-interfaces.

The Oxygen vacancies, which are developed at the interface during the deposition process, are considered as indispensable parts of the interface q2DEG and have very significant role in ferromagnetism, superconductivity and their coexistence~\cite{mohantaJPCM,NM_VacancyJPCM2014}. We model these non-magnetic impurities as the local random shifts in the chemical potential and express by the Hamiltonian ${\cal H}_{dis}=\sum_{i_d}^{N_d}\sum_{\alpha\sigma}V^{i_d} c_{i_d\alpha\sigma}^{\dagger}c_{i_d\alpha\sigma}$, where $i_d$ denotes the defect sites of total number $N_d$, $V^{i_d}$ is a random potential which varies within a range [$-W,W$] and $W$ is the strength of the disorder. The percentage defect concentration is given by $n_d=N_d/{N^2}\times100$.

The total effective Hamiltonian for the interface electrons is, therefore, given by
\begin{equation}
{\cal H}_{eff}={\cal H}_0+{\cal H}_{ASO}+{\cal H}_{RSO}+{\cal H}_{FM}+{\cal H}_{SC}+{\cal H}_{dis}
\label{H_eff}
\end{equation}
where ${\cal H}_0=\sum_{k,\alpha,\sigma}(\epsilon_{k\alpha}-\mu) c_{k\alpha\sigma}^\dagger c_{k\alpha\sigma}$ describes the band dispersion of the electrons in the three $t_{2g}$ orbitals with $\epsilon_{ka}=-2t_1(\cos{k_x}+\cos{k_y})-t_2-4t_3\cos{k_x}\cos{k_y}$, $\epsilon_{kb}=-t_1(1+2\cos{k_y})-2t_2\cos{k_x}-2t_3\cos{k_y}$, $\epsilon_{kc}=-t_1(1+2\cos{k_x})-2t_2\cos{k_y}-2t_3\cos{k_x}$, $\mu$ is the chemical potential and $t_1=0.277$ eV, $t_2=0.031$ eV, $t_3=0.076$ eV are the tight-binding parameters~\cite{Held_PRB2013}.

Although multi-band superconductivity in this interface q2DEG was proposed earlier~\cite{Caprara_PRB2013,Nakamura_JPSJ2013}, the explicit nature of the pairing symmetry and the intra-band or inter-band superconductivity were not explored. Our model uses the realistic band structures, obtained from DFT studies~\cite{Held_PRB2013}, and treats the electron pairing in intra-band and inter-band channels within the multi-band BdG theory, to be described below. 
 
\section{Multi-band BdG Theory}
\label{bdg}
The self-consistent BdG theory is perhaps the best available numerical technique to study the interplay of superconductivity with real-space inhomogeneity or competing orders such as ferromagnetism or charge density wave within mean-field approximation in any experimentally realizable geometry. To begin with, the Hamiltonian ${\cal H}_{eff}$ in Eq.~(\ref{H_eff}) is written in real lattice as
\begin{align}
{\cal H}_{BdG}&=-\sum_{ij,\alpha,\beta,\sigma,\sigma^{\prime}}(t_{\alpha\beta}^{ij\sigma\sigma^{\prime}}c_{i\alpha\sigma}^\dagger c_{j\beta\sigma^{\prime}}+h.c.)\nonumber\\
&-\sum_{i,\alpha,\sigma}(\mu-V^{i_d}\delta_{ii_d})c_{i\alpha\sigma}^{\dagger}c_{i\alpha\sigma}\nonumber\\
&+\sum_{i,\alpha,\sigma,\sigma^{\prime}}(h_{x\alpha}\sigma_x)_{\sigma \sigma^{\prime}}c_{i\alpha\sigma}^\dagger c_{i\alpha\sigma^{\prime}}\nonumber\\
&+\sum_{i,\alpha,\beta}(\Delta_{\alpha\beta}^{s}(r_i)c_{i\alpha\uparrow}^{\dagger}c_{i\beta\downarrow}^{\dagger}+h.c.)\nonumber\\
&+\sum_{<ij>,\alpha,\beta,\sigma}(\Delta_{\alpha\beta\sigma\sigma}^{t}(r_i)c_{i\alpha\sigma}^{\dagger}c_{j\beta\sigma}^{\dagger}+h.c.)
\label{H_bdg}
\end{align}
where $t_{\alpha\beta}^{ij\sigma\sigma^{\prime}}$ is the tight-binding hopping amplitudes which contains ${\cal H}_0$, ${\cal H}_{ASO}$ and ${\cal H}_{RSO}$, $\Delta_{\alpha\beta}^{si}=-g\braket{c_{i\alpha\uparrow}c_{i\beta\downarrow}}$ and $\Delta_{\alpha\beta\sigma\sigma}^{ti}=-g\braket{c_{i\alpha\sigma}c_{i\beta\sigma}}$ are the local singlet and triplet pairing gaps, $\delta_{ii_d}$ is the Kronecker's delta function.

The Hamiltonian ${\cal H}_{BdG}$ in Eq.~(\ref{H_bdg}) is diagonalized by the unitary Bogoliubov transformation $\hat{c}_{i\alpha\sigma}=\sum_{n,\sigma^{\prime}}u_{n\alpha\sigma}^{i\sigma^{\prime}}\hat{\gamma}_{n}^{\sigma^{\prime}}+v_{n\alpha\sigma}^{i\sigma^{\prime}*}\hat{\gamma}_{n}^{\sigma^{\prime}\dagger}$ which yields the multi-band BdG equations ($\sigma^{\prime}$, being a dummy index, is omitted hereafter.)
\begin{align}
&\sum_j \begin{pmatrix} \begin{array}{cccc} \sum_{\beta}\Gamma_{\alpha\beta}^{ij\uparrow\uparrow} & \sum_{\beta}\Gamma_{\alpha\beta}^{ij\uparrow\downarrow} & \Delta_{\alpha\beta\uparrow\uparrow}^{tij}\delta_{\alpha\beta} & \Delta_{\alpha\beta}^{sij}\delta_{\alpha\beta} \\ \sum_{\beta}\Gamma_{\alpha\beta}^{ij\downarrow\uparrow} & \sum_{\beta}\Gamma_{\alpha\beta}^{ij\downarrow\downarrow} & -\Delta_{\alpha\beta}^{sij}\delta_{\alpha\beta} & \Delta_{\alpha\beta\downarrow\downarrow}^{tij}\delta_{\alpha\beta} \\ \Delta_{\alpha\beta\uparrow\uparrow}^{tij*}\delta_{\alpha\beta} & -\Delta_{\alpha\beta}^{sij*}\delta_{\alpha\beta} & -\sum_{\beta}\Gamma_{\alpha\beta}^{ij\uparrow\uparrow*} & -\sum_{\beta}\Gamma_{\alpha\beta}^{ij\uparrow\downarrow *}\\\Delta_{\alpha\beta}^{sij*}\delta_{\alpha\beta} & \Delta_{\alpha\beta\downarrow\downarrow}^{tij*}\delta_{\alpha\beta} & -\sum_{\beta}\Gamma_{\alpha\beta}^{ij\downarrow\uparrow *} & -\sum_{\beta}\Gamma_{\alpha\beta}^{ij\downarrow\downarrow *}\end{array} \end{pmatrix}\nonumber\\
&\times\begin{pmatrix} \begin{array}{c} u_{n\alpha\uparrow}^j \\ u_{n\alpha\downarrow}^j \\ v_{n\alpha\uparrow}^j \\ v_{n\alpha\downarrow}^j \end{array} \end{pmatrix}=E_n \begin{pmatrix} \begin{array}{c} u_{n\alpha\uparrow}^j \\ u_{n\alpha\downarrow}^j \\ v_{n\alpha\uparrow}^j \\ v_{n\alpha\downarrow}^j \end{array} \end{pmatrix}
\label{EBdG}
\end{align}
where $\Gamma_{\alpha\beta}^{ij\sigma\sigma^{\prime}}=- t_{\alpha\beta}^{ij\sigma\sigma^{\prime}}-[(\mu-V^{i_d}\delta_{ii_d})\delta_{\sigma\sigma^{\prime}}-(h_{x\alpha}\sigma_x)_{\sigma\sigma^{\prime}}]\delta_{ij}\delta_{\alpha\beta}$.
For a square lattice of size $N \times N$, the BdG Hamiltonian matrix has dimension $12N^2 \times 12N^2$ for three orbitals. Using the above Bogoliubov transformation, the local pairing gaps can be obtained as 
\begin{align}
&\Delta_{\alpha\beta}^{si}=-\frac{g}{2}\sum_{n}[u_{n\alpha\uparrow}^i v^{i*}_{n\beta\downarrow}-u_{n\alpha\downarrow}^i v^{i*}_{n\beta\uparrow}]\tanh{\left(\frac{E_n}{2k_BT}\right)}\nonumber\\
&\Delta_{\alpha\beta\sigma\sigma}^{ti}=-\frac{g}{2}\sum_{n,<j>}[u_{n\alpha\sigma}^i v^{j*}_{n\beta\sigma}-u_{n\alpha\sigma}^i v^{j*}_{n\beta\sigma}]\tanh{\left(\frac{E_n}{2k_BT}\right)}
\label{gap_eq}
\end{align}
where $k_B$ is the Boltzmann constant and $T$ is the temperature. The total occupation number $n=(1/N^2)\sum_{i,\alpha,\sigma}\braket{c_{i\alpha\sigma}^{\dagger}c_{i\alpha\sigma}}$ is computed using the following relation
\begin{equation}
n=\frac{1}{N^2}\sum_{n,i,\alpha,\sigma}[|u_{n\alpha\sigma}^i|^2 f(E_n)+|v_{n\alpha\sigma}^i|^2 (1-f(E_n))]
\label{occup}
\end{equation}
where $f(E_n)=1/(1+\exp(E_n/k_{B}T))$ is the fermi function. In what follows, the BdG equations~(\ref{EBdG}) are solved numerically on a square lattice with periodic boundary conditions to find out the eigenvalues $E_n$ and the local quasi-particle amplitudes $u_{n\alpha\sigma}^i$, $u_{n\alpha\sigma}^i$ and the new pairing gaps are calculated using Eq.~(\ref{gap_eq}). This process is repeated until self-consistency is reached at every lattice sites. Finally, the average values are obtained via $\Delta_{\alpha\beta}^{s/t}=(1/N)\sum_i\Delta_{\alpha\beta}^{(s/t)i}$.

It is important to mention here that static mean-field theory neglects fluctuations and, therefore, over-estimates the fermionic field amplitudes. The temperature or the magnetic field, being treated in the analysis, are therefore shouldbe seen as just parameters and their qualitative features, not quantitative estimates, are relevant.
\section{Results}
\label{results}
Having formulated the self-consistent BdG theory, we now present the numerical results obtained on a $25\times25$ square lattice. First we analyze, in the homogeneous situation ($n_d=0$), the effects of Rashba and atomic SOI on superconductivity, the gap-structures and phase diagrams of the singlet and triplet pairings and then the effects of in-plane magnetic field on superconductivity.

Spin orbit interactions are known to have unusual effects on superconductivity~\cite{Kim_PRB2012,Sigrist_EPL2000}.  
\begin{figure}[htb!]\vspace{0em}
\begin{center}
\epsfig{file=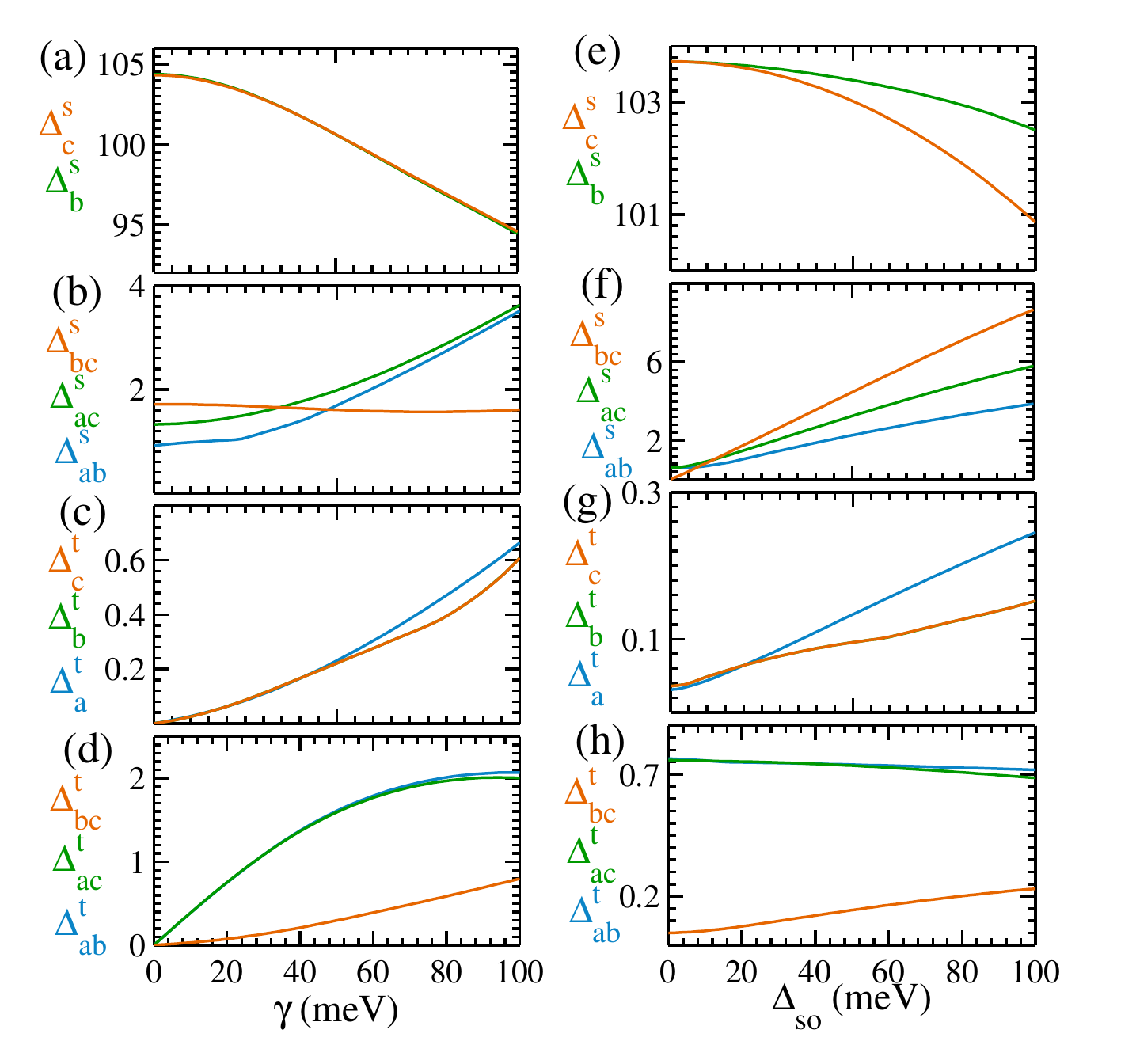,trim=0.05in 0.15in 0in 0.1in,clip=true, width=90mm}\vspace{0em}
\caption{(Color online)  The variation of the singlet and triplet pairing amplitudes (in meV), in intra-band and inter-band channels, with respect to the Rashba and atomic SOI strengths $\gamma$ (left column) and $\Delta_{so}$ (right column) respectively. Other parameters are $\mu=-0.6$ eV, $T=0$, $g=0.135$ eV, $h_{x1}=0.4$ eV, $h_{x2}=0.1$ eV and $n_d=0$.}
\label{gma_var}\vspace{0em}
\end{center}
\end{figure}
The gate-tunable Rashba SOI makes the interface q2DEG a potential candidate for spintronic applications~\cite{Bibes_PTRSA2012} as well as a playground for the search of non-trivial topological excitation such as the Majorana fermions~\cite{Mohanta_EPL2014} or the Skyrmions~\cite{Leon_PRB2014,Garaud_PRB2014} or novel Hall phases~\cite{mohanta_ahesc}. We study the effects of Rashba SOI and atomic SOI on the intra-band and inter-band pairing as shown in FIG.~\ref{gma_var}. The pairing gap $\Delta_{a}$ in the $d_{xy}$ orbital is largely suppressed by the in-plane ferromagnetism while the pairing gaps $\Delta_{b}$ and $\Delta_{c}$ in the $d_{yz}$, $d_{zx}$ orbitals dominate. As depicted in FIG.~\ref{gma_var}(a), (e), the pairing amplitudes $\Delta_{b}$ and $\Delta_{c}$ decreases slowly with increasing both Rashba SOI strength $\gamma$ and atomic SOI strength $\Delta_{so}$ because the SOI enhances precession of electrons leading to slow reduction of the electron pairing. It is interesting to note that the superconductivity in the singlet inter-band channel and the triplet intra-band and inter-band channels is induced by the SOI when the pairing amplitudes $\Delta_b$ or $\Delta_c$ is finite as shown in FIG.~\ref{gma_var}(b)-(d),(f)-(h). This is because of the fact that the SOI breaks the spin-degeneracy and, in presence of the Zeeman splitting, the spin and orbitals nature of the electronic bands are re-organized and the pairings, in these channels, become energetically favourable. 

\begin{figure}[t]\vspace{0em}
\begin{center}
\epsfig{file=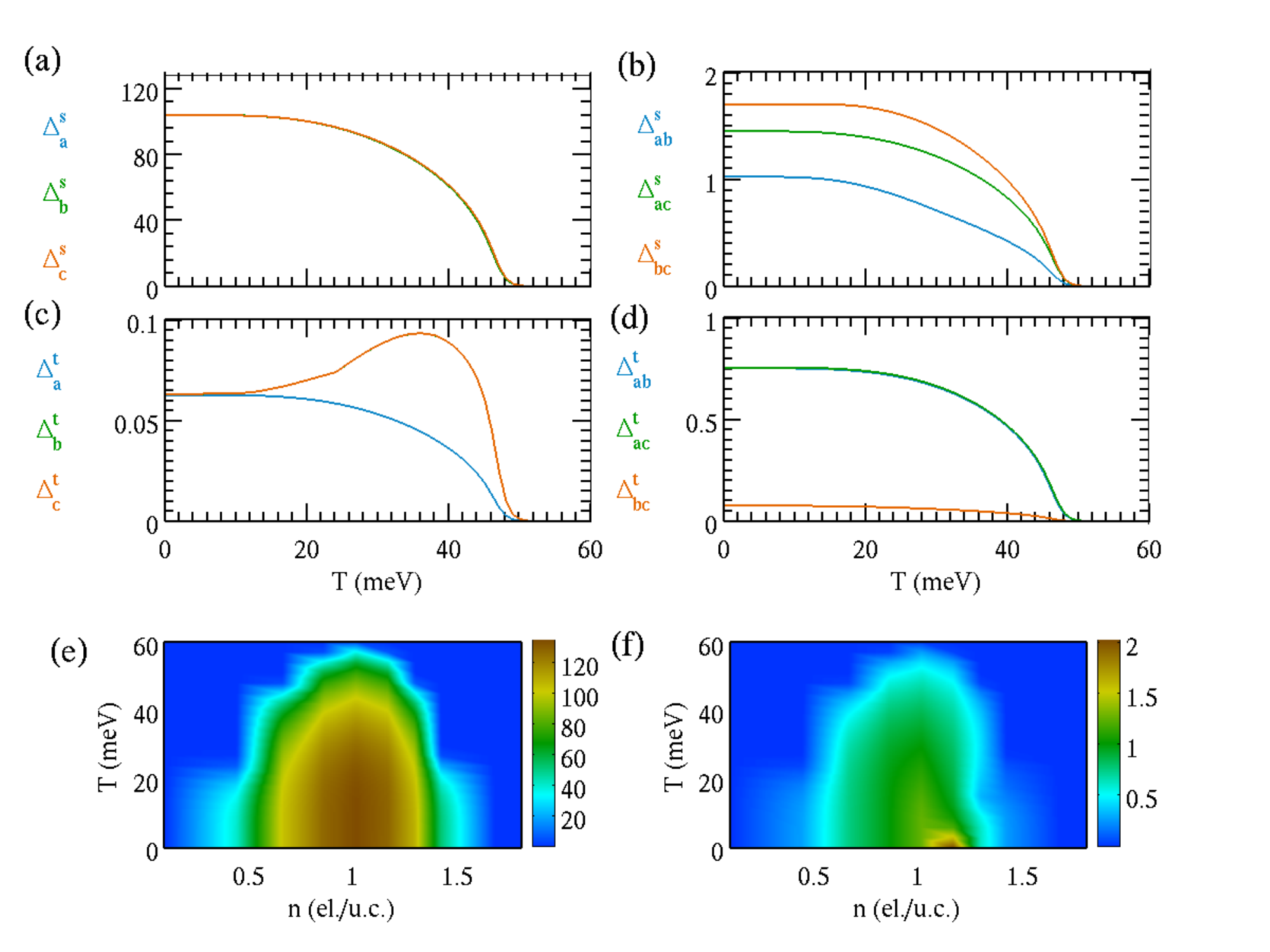,trim=0.1in 0.05in 0.2in 0.28in,clip=true, width=90mm}\vspace{0em}
\caption{(Color online)  (a)-(d) The gap structures of the singlet and triplet superconductivity, in intra-band and inter-band channels. All pairing amplitudes are in meV. The maximum (e) singlet and (f) triplet pairing amplitudes are plotted in the $n-T$ plane. Other parameters are $g=0.135$ eV, $h_{x1}=0.4$ eV, $h_{x2}=0.1$ eV and $n_d=0$.}
\label{mu_T_del}\vspace{0em}
\end{center}
\end{figure}

The temperature-variations of the pairing amplitudes are shown in FIG.~\ref{mu_T_del}(a)-(d). The pairing gaps reveal BCS nature in all the pairing channels except the triplet intra-band pairing $\Delta_c^t$ which gets enhanced near the transition temperature.
We plot the maximum pairing amplitudes in the singlet and triplet channels in the $n-T$ phase-plane as described in FIG.~\ref{mu_T_del}(e)-(f). which show the superconducting phases. It is evident that both the singlet and triplet pairing channels show a dome-shaped superconducting phase and the triplet pairing becomes stronger towards higher values of career density. 

\begin{figure}[t]\vspace{0em}
\begin{center}
\epsfig{file=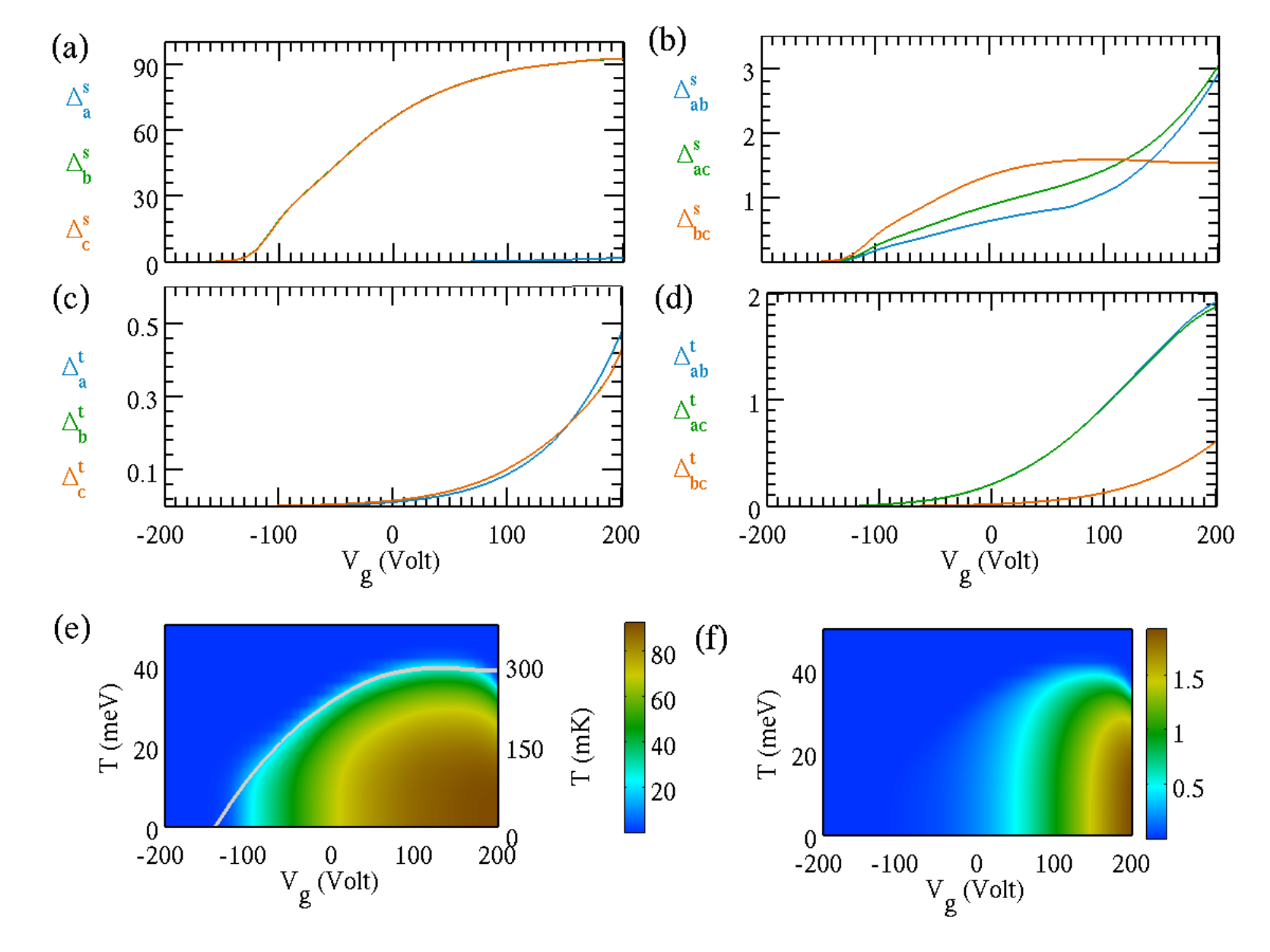,trim=0.3in 0.15in 0.25in 0.15in,clip=true, width=87mm}\vspace{0em}
\caption{(Color online) (a)-(d) Variation of the pairing amplitudes with respect to external gate voltage $V_g$. The maximum (e) singlet and (t) triplet pairing amplitudes are plotted in the $V_g-T$ plane. The white line in figure (e) is the experimental data, obtained from Ref.~\onlinecite{Caviglia2008}, showing the variation of the transition temperature $T_c$ with $V_g$. Other parameters are $g=0.135$ eV, $h_{x1}=0.4$ eV, $h_{x2}=0.1$ eV and $n_d=0$.}
\label{vg_var}\vspace{0em}
\end{center}
\end{figure}

In the experiments, a gate voltage ($V_g$) is tuned to control the doping level of the q2DEG and a superconductor to insulator transition with a quantum critical point $V_g=140$ V~\cite{Caviglia2008} is evinced by varying the gate voltage. It is, therefore, fascinating to study the variation of the pairing amplitudes in different channels with respect to $V_g$ and eventually to plot the phase-diagram in the $V_g-T$ plane. In appendix~\ref{gate}, we derive the gate-voltage dependence of the career density and Rashba SOI. In FIG.~\ref{vg_var}(a)-(d), we plot the gate-voltage variation of the pairing amplitudes which reveal the superconducting transition at $V_g\simeq 140$ V. It is interesting to note that the singlet intra-band pairing gaps and triplet intra-band and inter-band pairing gaps follow the nature of the variation of the Rashba SOI with respect to $V_g$ as shown in Fig.~\ref{gate_var}(b). 
FIG.~\ref{vg_var}(e)-(f) depict the superconducting phases in the singlet and triplet channels respectively. The superconducting phase shown in FIG.~\ref{vg_var}(e) fits well with the experimental data except a quantitative mismatch of the transition temperatures due to inherent overestimation problem of mean-field theory. The triplet pairing appears to begin at higher values in the gate voltage range. It has been found experimentally that the pairing gap $\Delta$ behaves differently than the transition temperature $T_c$ with respect to $V_g$~\cite{RichterNature2013}. In the underdoped region, $T_c$ increases with $V_g$ while $\Delta$ decreases. This unusually different variation of $\Delta$ and $T_c$ with $V_g$ has been elucidated as due to the precise nature of the phonon spectral function $\alpha^2g(\omega)$ in the underdoped regime and due to the entrance of new bands which suppress $\Delta$ in the overdoped regime~\cite{Boschker_arXiv2015}.

\begin{figure}[t]\vspace{0em}
\begin{center}
\epsfig{file=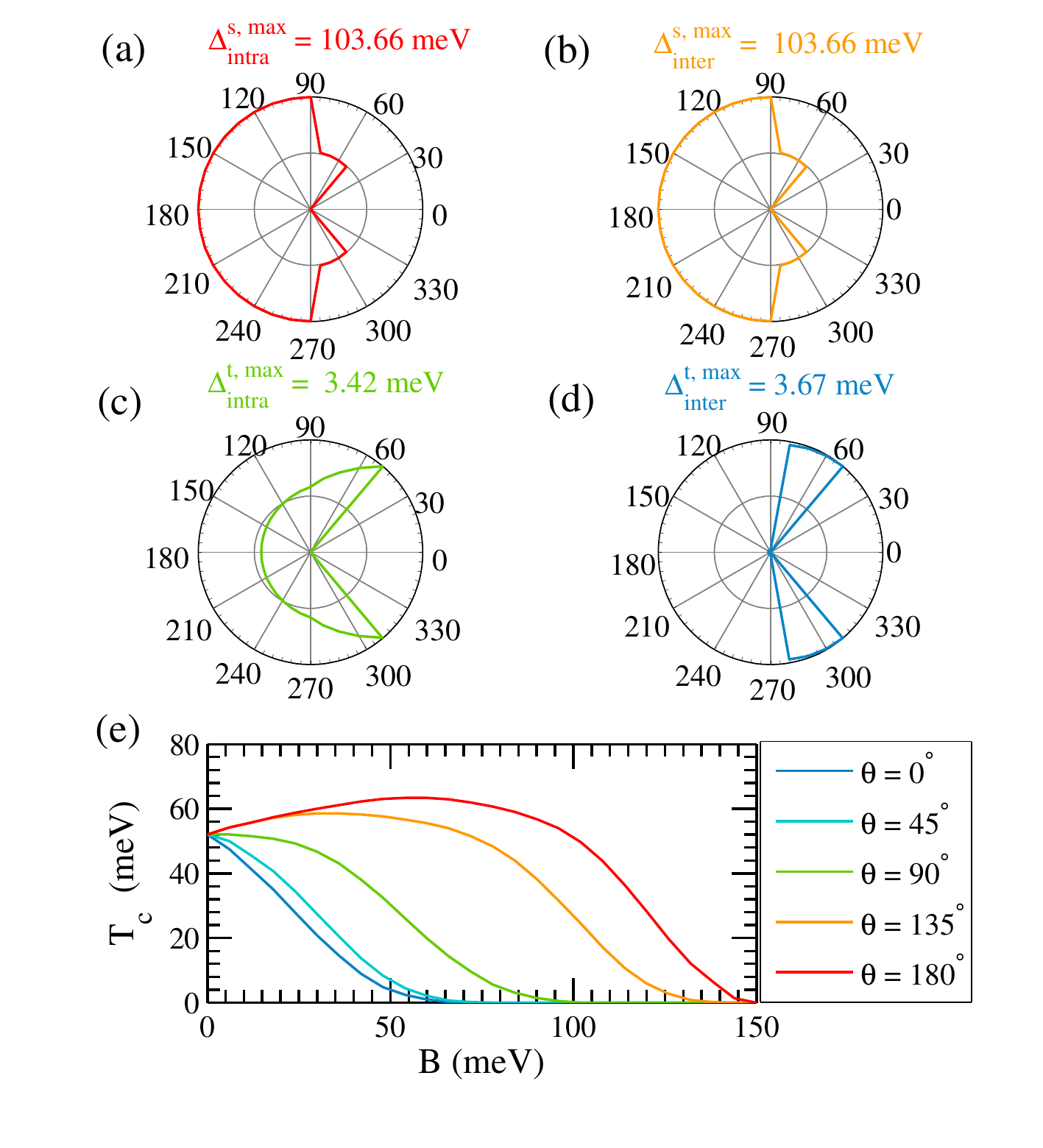,trim=0.5in 0.4in 0.1in 0.in,clip=true, width=90mm}\vspace{0em}
\caption{(Color online) The polar plots in figures (a)-(d) show  the angular-variations of the maximum of singlet and triplet pairing amplitudes in the intra-band and inter-band channels respectively. The angle $\theta$ is between the initial polarization direction due to the intrinsic ferromagnetism and the applied magnetic field. Figure (e) displays the variation of the superconducting transition temperature $T_c$ with respect to the amplitude $B$ of the magnetic field. Parameters used are $\mu=-0.6$ eV, $T=0$ eV, $g=0.135$ eV, $h_{x1}=0.4$ eV, $h_{x2}=0.1$ eV and $n_d=0$.}
\label{theta_var}\vspace{0em}
\end{center}
\end{figure}

Next we study the effect of external in-plane magnetic field on superconductivity. As shown in FIG.~\ref{theta_var}, the magnetic field is applied in the interface plane along different directions ($\theta$) with respect to the initial direction of polarization due to in-plane ferromagnetism and the pairing gaps are plotted in the polar plane of ($\Delta$, $\theta$). 
The singlet pairing amplitudes are  increased along the opposite direction of that of the intrinsic polarization direction while the triplet pairing amplitudes get strengthened along $\theta\simeq 60^{\circ},300^{\circ}$. FIG.~\ref{theta_var}(e) plots the superconducting transition temperature $T_c$ as a function of the amplitude $B$ of the applied magnetic field. Remarkably, the enhancement of superconductivity by in-plane magnetic field has been reported experimentally in LAO/STO interface~\cite{Gardner2011}. The enhancement of superconductivity is because of the interplay between superconductivity, the intrinsic ferromagnetism and the ferromagnetism induced by the applied magnetic field and implicates a possibility of a hidden superconducting phase above the transition temperature. The hidden superconducting phase is also inferred from the recent observation of the magnetic field assisted transient superconducting state observed in the interface q2DEG at 245 mK~\cite{Aveek_arxiv2014}. It is relevant to mention here that a hidden order has been proposed to be the precursor of superconductivity in Fe-pnictdes and high-Tc Cuprates~\cite{Moor_PRB2015}. It may, as well, be possible that a hidden superconducting phase, arising from the competition between ferromagnetism and superconductivity, exist in LAO/STO hetero-interface.

\section{Monte-Carlo study of thermal phase-fluctuation}
\label{mc_phase}
The interface q2DEG is highly inhomogenous in nature and becomes superconducting at very low career concentrations. Therefore, the superconductivity is prone to the detrimental effects of non-magnetic disorder and the role of phase-fluctuation becomes important. In a previous study~\cite{mohantaJPCM}, we show that disorder, in fact, can help the antagonistic ferromagnetism to live apart with superconductivity in spatially phase-segregated regions at the interface. 
\begin{figure}[t]\vspace{0em}
\begin{center}
\epsfig{file=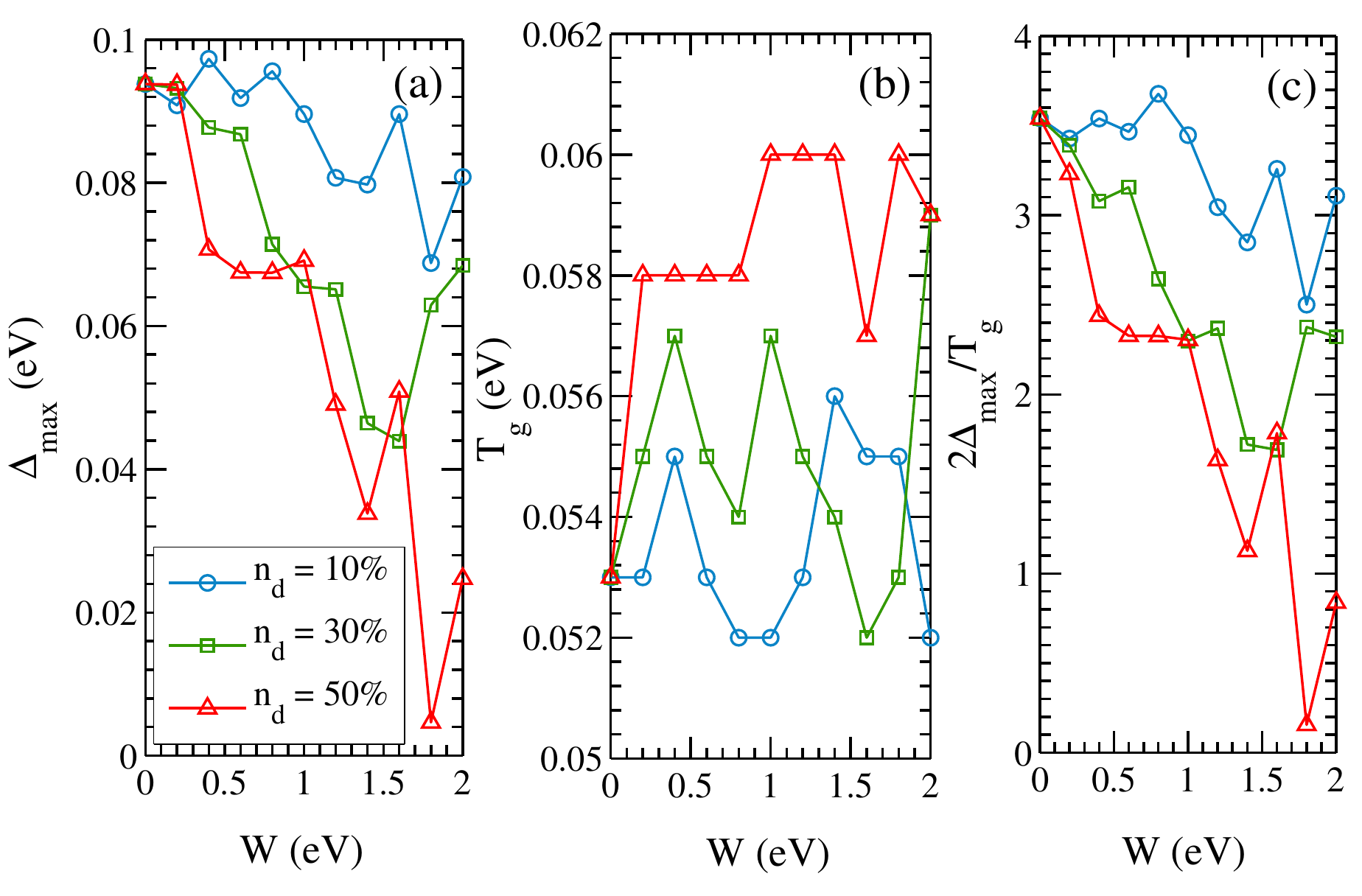,trim=0in 0in 0.in 0.in,clip=true, width=85mm}\vspace{0em}
\caption{(Color online)  The variation of (a) the maximum pairing amplitude $\Delta_{max}$, (b) the gap-closing temperature $T_g$ and (c) the BCS ratio $2\Delta_{max}/T_{g}$ with respect to the disorder strength $W$ for different disorder concentrations $n_d$. Other parameters are $\mu=-0.6$ eV, $g=0.135$ eV, $h_{x1}=0.4$ eV and $h_{x2}=0.1$ eV.}
\label{W_var}\vspace{0em}
\end{center}
\end{figure}
In FIG.~\ref{W_var}, we show the variation of the maximum pairing gap $\Delta_{max}$, the gap-closing temperature $T_g$ and the BCS ratio $2\Delta_{max}/T_{g}$ with respect to disorder strength $W$ for different concentrations $n_d$. With increasing $W$ and $n_d$, both $\Delta_{max}$ and the BCS ratio decreases with fluctuation. On the other hand, $T_g$ fluctuates, with increasing $W$, within a narrow range of temperature and shows increasing tendency with increasing $n_d$. It is evident that in the highly disorder limit, $\Delta$ and $T_g$ behave differently and, therefore, can no longer track the superconducting transition. It is required to go beyond the standard BCS prescription and incorporate the phase-fluctuation into account in the effective theory.

We use a Monte Carlo (MC) technique in conjunction with self-consistent BdG formalism to study the thermal phase-fluctuations near the superconducting transition driven by magnetic field or temperature. In the MC analysis, we consider only singlet intraband pairings in the $d_{yz}$ and $d_{zx}$ orbitals as they dominate over the pairings in other channels substantially and use a single pairing gap $\Delta=\Delta_b^s\simeq\Delta_c^s$ to reduce the computation time. The complex superconducting order parameter is taken as $\Delta(r_i)=|\Delta_i|e^{j\phi_i}$, where $|\Delta_i|$ and $\phi_i$ are, respectively, the amplitude and phase of the pairing gap at site $i$. We start at high temperature with a specific disorder-configuration and reach low temperature upto 0.002 eV. At each temperature, the site-resolved pairing gaps $|\Delta_i|$ are computed using the self-consistent BdG method, described in section~\ref{bdg}, and are fed into the MC update process which uses a free-energy minimization technique to determine the phases $\phi_i$. In appendix~\ref{free_en}, we present the calculation of the free energy for a given BdG Hamiltonian.  
\begin{figure}[t]\vspace{0em}
\begin{center}
\epsfig{file=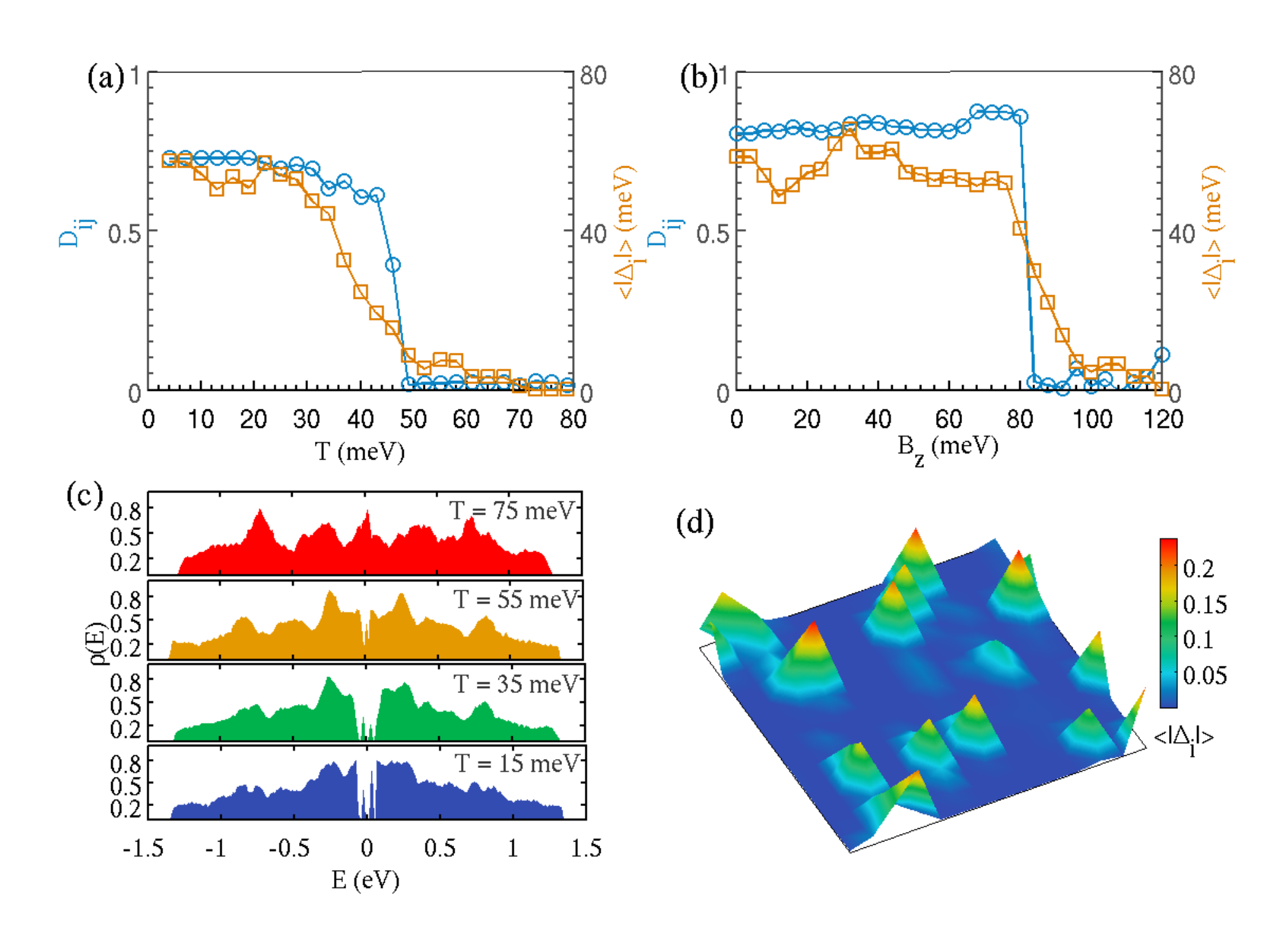,trim=0.25in 0.35in 0.in 0.1in,clip=true, width=90mm}\vspace{0em}
\caption{(Color online)  The variation of the phase correlation function $D_{ij}$ and the disordered and site-averaged pairing amplitude $\braket{|\Delta_i|}$ with respect to (a) temperature $T$ and (b) perpendicular magnetic field $B_z$. Figure (c) depicts the density of states $\rho(E)$ at different temperatures across the transition to the superconducting state. (d) The profile of the pairing amplitude $\braket{|\Delta_i|}$ at $T=60$ meV describing the localized Cooper pairs in the non-superconducting side of the transition (in figure (a)). A $14\times14$ lattice is used in the calculation. Other parameters are $\mu=-0.6$ eV, $g=0.135$ eV, $h_{x1}=0.4$ eV, $h_{x2}=0.1$ eV, $W=0.8$ eV and $n_d=50$.}
\label{phase_fluc}\vspace{0em}
\end{center}
\end{figure}

In the weak to moderate disorder limit, the pairing gap parameter tracks the superconducting transition. However, in the highly disordered limit, in which the interface q2DEG lies, the superconducting transition is indicated by the destruction of the global phase-coherence. To quantify the phase-coherence, we compute a long-ranged phase correlation function $D_{ij}=\braket{\cos(\phi_i-\phi_j)}$, similar to that used in the XY-model to track the universal BKT transition, where $i$ and $j$ are sites separated by large distance and $`\braket{\ }$' denotes the disorder and site averaged value. In FIG.~\ref{phase_fluc}(a)-(b), we show the variation of $D_{ij}$ and $\braket{|\Delta_i|}$ with respect to temperature $T$ and perpendicular magnetic field $B_z$ respectively. In both the cases, the transition to the non-superconducting phase is dictated by a vanishing $D_{ij}$ and, on the other hand, $\braket{|\Delta_i|}$ remains finite and fluctuates. The small finite value of $\braket{|\Delta_i|}$ in the non-superconducting phase is the signature of localized Cooper pairs~\cite{dubinature2007}. The profile of the pairing gap in the two-dimensional space is shown in FIG.~\ref{phase_fluc}(d). The direct evidence of the presence of localized Cooper pairs in the insulating phase has been found in recent transport measurement~\cite{Mehta2012}. FIG.~\ref{phase_fluc}(c) shows the single particle density of states, given by
\begin{equation*}
\rho(E)=\frac{1}{N}\sum_{n,i,\alpha,\sigma}\left[{(u_{n\alpha\sigma}^{i})}^2\delta(E-E_n)+{(v_{n\alpha\sigma}^{i})}^2\delta(E+E_n) \right],
\end{equation*}
at different temperatures across the superconducting transition. It is important to note that because of strong disorder, quasi-particle states, bound to the impurities, appear within the bulk superconducting gap and the system passes through a pseudo-gap like phase during the quantum phase transition~\cite{RichterNature2013}.

\section{Conclusions}
\label{conclusions}
In this paper, we presented a three-band model for the superconductivity at the LaAlO$_3$/SrTiO$_3$ (001) interface. We explored, using a multi-orbital BdG theory, the interplay between the superconductivity, ferromagnetism, and spin-orbit interactions. We also studied the role of thermal phase fluctuation using a Monte-Carlo method. The key findings of our analysis can be summarized as follows.

The electron pairing in the $d_{xy}$ band is suppressed by the competing ferromagnetic order and the pairing in the $d_{yz}$, $d_{zx}$ bands dominates. We find that the singlet pairing in the inter-band channel and the triplet pairing in both intra-  and inter-band channel is induced by the SOI when the pairing amplitude in the singlet intra-band channel is finite. 

Gate-voltage has been found to affect supercocinductivity at the interface quite strongly. We calculated the gate-voltage variation of the pairing amplitudes in different channels and  extracted the singlet and triplet superconducting phases in the $V_g-T$ plane and compared with experimental results with a fair degree of agreement. 

We observed an enhancement of superconductivity by external in-plane magnetic field. We found that the superconducting transition temperature depends on the direction of the applied in-plane magnetic field which is an expermentally verifiable prediction from our theory. The enhancement of superconductivity by applied magnetic field is due to the interplay between ferromagnetism and superconductivity and suggests a 'hidden superconducting phase' above the transition temperature. The recent observation of magnetic field-assisted transient superconducting state at $245$ mK~\cite{Aveek_arxiv2014} agrees with this hidden superconducting order and needs to be explored further experimentally~\cite{NM_AT_tobe}.

Lastly, we studied the role of thermal phase fluctuation on the superconductivity using a Monte-Carlo method and found that there exist localized Cooper pairs in the non-superconducting phase beyond the quantum phase transition driven by perpendicular magnetic field. The same is also seen in the normal state just above the superconducting transition temperature. The density of states reveal that, in the highly disordered situation, quasi-particle bound states appear within the bulk superconducting gap and the system passes through a pseudo-gap like phase exactly as reported in the tunneling spectroscopic measurement~\cite{RichterNature2013}.

\section*{Acknowledgments}
We thank Aveek Bid and Siddhartha Lal for fruitful discussions and acknowledge the use of the computing facility from DST-FIST (phase-II) Project installed in the Department of
Physics, IIT Kharagpur, India. 

\hspace{2em}
\appendix
\section{Gate-voltage dependence}
\label{gate}
The voltage $V_g$, applied by back-gating to the STO substrate with the interface 2DEG grounded, controls the career density at the interface and the Rashba spin-orbit splitting~\cite{Caviglia2008, CavigliaPRL2010}. The career density $n$ changes with the applied electric field $F$ according to the relation
\begin{equation}
n(F)=\frac{2\epsilon_0}{e} \int_0^F \epsilon_r(F) dF
\end{equation}
where $\epsilon_0$ is the free-space permittivity, $\epsilon_r(F)$ is the field-dependent relative permittivity and $e$ is the quantum of electronic charge. For bulk STO, $\epsilon_r(F)=1/A(1+\frac{B}{A}F)$, where $A=4.097\times10^{-5}$ and $B=4.907\times10^{-10}$ are temperature-dependent parameters, determined experimentally at 4.3 K by a first-order fit to the permittivity~\cite{Neville_JAP1972}. However, it has a singularity at $F=-A/B$ which is circumvented, as prescribed in Ref.~\onlinecite{Balatsky_PRB2012}, by taking into account the second order term in the denominator. The corrected expression is given by
$\epsilon_r(F)=1/A[C_1+C_2\frac{B}{A}F+C_3(\frac{B}{A})^2F^2]$,
where $C_i$ ($i=1,2,3$) are parameters to be found out by experimental fits.  
\begin{figure}[htb!]\vspace{0em}
\begin{center}
\epsfig{file=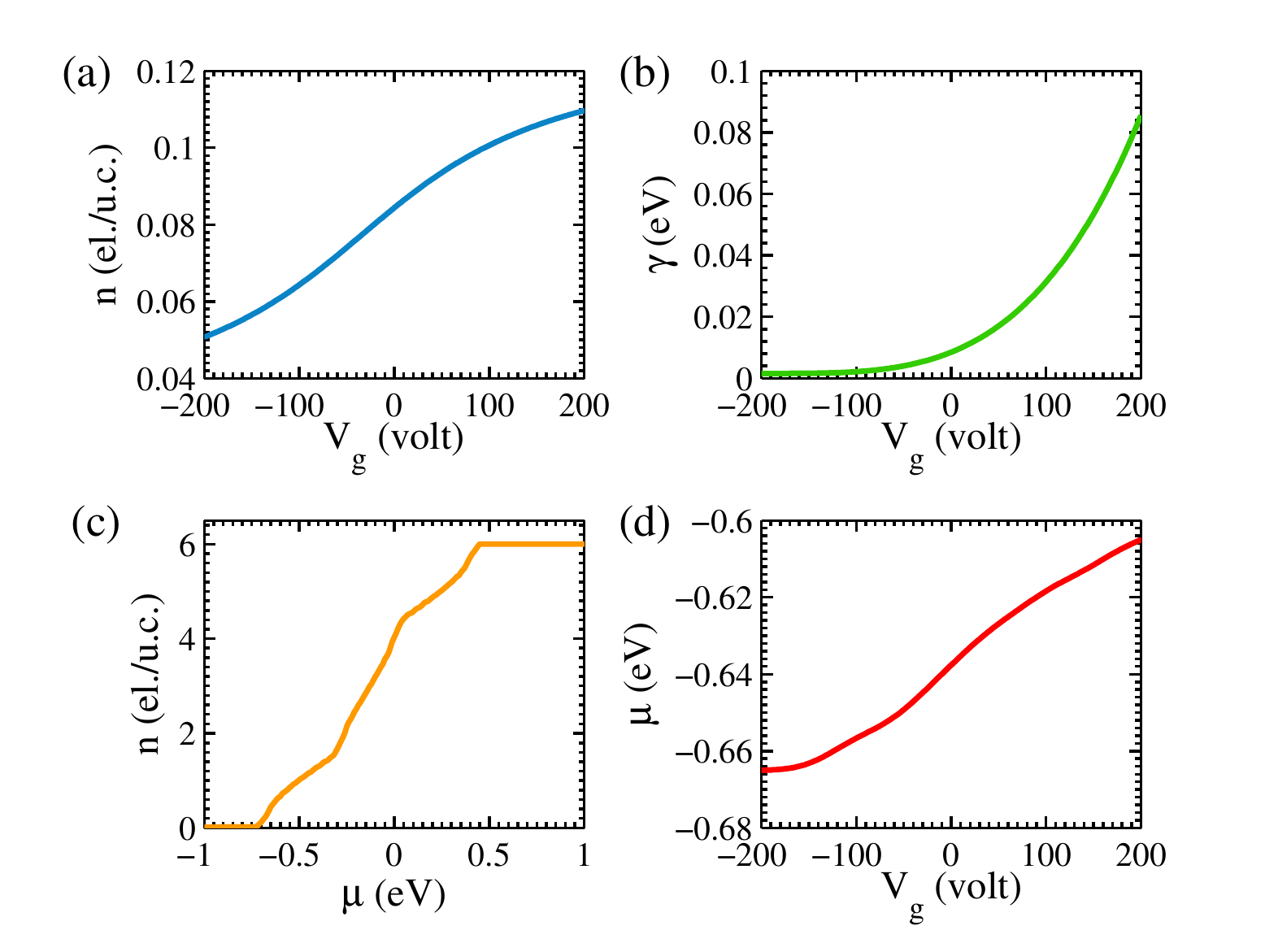,trim=0.1in 0.05in 0.in 0.1in,clip=true, width=90mm}\vspace{0em}
\caption{(Color online)  Gate voltage modulation of (a) the career concentration and (b) the strength of Rashba spin-orbit splitting fitted with the experimental data~\cite{Caviglia2008, CavigliaPRL2010}. The variation of (c) the total occupation number as a function of the chemical potential and (d) the chemical potential as a function of the gate voltage at zero temperature. A career concentration of $10^{13}$ cm$^{-2}$ is equivalent to $0.016$ el./u.c~\cite{Bucheli_PRB2014}.}
\label{gate_var}\vspace{0em}
\end{center}
\end{figure}
The career density can, therefore, be expressed as 
\begin{equation}
n(F)=\frac{4\epsilon_0\Big\{ \tan^{-1} \Big[ \frac{2C_3BF+C_2A}{A\sqrt{C_4}}\Big] -\tan^{-1}\Big(\frac{C_2}{\sqrt{C_4}}\Big) \Big\}}{eB\sqrt{C_4}}
\end{equation}
where $C_4=4C_1C_3-C_2^2$. The gate-voltage $V_g$ enters in the above expression via the relation $F=V_g/d+F_0$, where $d=0.5$ mm is the thickness of the STO layer and $F_0=1.2\times10^5$ V/m is the initial electric field at the interface in the absence of the gate-voltage. Comparison with the experimental data~\cite{Caviglia2008} yields $C_1=4.25$, $C_2=-0.37$ and $C_3=0.29$. Fig.~\ref{gate_var}(a) describes the gate-voltage modulation of the career density.

Furthermore, the gate voltage tunes the Rashba spin-orbit splitting at the interface~\cite{CavigliaPRL2010}. In Ref.~\onlinecite{Bucheli_PRB2014}, the gate-voltage dependence of the Rashba spin-orbit interaction is derived using a Kane $\mathbf {k \cdot p}$ approach. However, it is interesting to note that the DFT studies in Ref.~\onlinecite{Held_PRB2013} reveals that the Rashba spin-orbit interaction at the interface is very different from usual Rashba type spin-orbit interaction found in semiconductor hetero-interfaces. Here, we fit the gate-voltage dependence of the Rashba splitting amplitude observed in Ref.~\onlinecite{CavigliaPRL2010}, to the equation $\gamma=a+b(V_g+200)^{2c}$ within $V_g$-range [$-200 ,200$] V, with $a=0.0016$, $b=3.6\times10^{-11}$ and $c=1.8$ as plotted in Fig.~\ref{gate_var}(b).

\hspace{2em}
\section{Free energy of BdG Hamiltonian}
\label{free_en}
The free energy of an inhomogenous superconductor is given by~\cite{Leggett_PRB98} 
\begin{equation}
{\cal F}=\braket{{\cal H}_{eff}}-T{\cal S}
\label{free}
\end{equation}
where $\braket{{\cal H}_{eff}}$ is the effective Hamiltonian of the superconductor, $T$ is the temperature and ${\cal S}$ is the entropy.

The effective mean-field Hamiltonian of a single-orbital system is written as 
\begin{equation}
{\cal H}_{eff}=\sum_{ij,\sigma}H_{ij} c_{i\sigma}^{\dagger}c_{i\sigma}+\sum_{i}[\Delta_i c_{i\uparrow}^{\dagger}c_{i\downarrow}^{\dagger}+h.c.]+\frac{|\Delta_i|^2}{U}
\end{equation} 
where $H_{ij}$ is the tight-binding Hamiltonian containing kinetic energy terms, $\Delta_i$ is the complex superconducting pairing gap at site $i$ and $U$ is the strength of the attractive electron-electron interaction.

The effective Hamiltonian ${\cal H}_{eff}$ is diagonalized by using the Bogoliubov transformation 
$\hat{c}_{i\sigma}=\sum_{n,\sigma^{\prime}}u_{n\sigma\sigma^{\prime}}^{i}\hat{\gamma}_{n\sigma^{\prime}}+v_{n\sigma\sigma^{\prime}}^{i*}\hat{\gamma}_{n\sigma^{\prime}}^{\dagger}$ to obtain
\begin{equation}
{\cal H}_{eff}=E_g+\sum_{n}E_n \gamma_{n\sigma^{\prime}}^{\dagger}\gamma_{n\sigma^{\prime}}
\end{equation}
where $E_n$ are the eigen-values of the BdG equations and the ground state energy $E_g$ is given by
\begin{equation}
E_g=-\sum_{i,n,\sigma}E_n |v_{n\sigma}^i|^2+\sum_{i}\frac{|\Delta_i|^2}{U}
\end{equation}

The entropy of an ideal gas of fermionic quasi-particles is expressed as~\cite{Landau_Lifshitz}
\begin{equation}
{\cal S}=-\sum_{n}[f_n \ln f_n+(1-f_n) \ln (1-f_n)]
\end{equation}

Therefore, using $\braket{\gamma_{n\sigma^{\prime}}^{\dagger}\gamma_{n\sigma^{\prime}}}=f_{n\sigma^{\prime}}$ and $f_{n\uparrow}=f_{n\downarrow}=f_{n}$, Eq.~(\ref{free}) becomes
\begin{align}
{\cal F}&=-\sum_{i,n,\sigma}E_n |v_{n\sigma}^i|^2+\sum_{i}\frac{|\Delta_i|^2}{U}+\sum_{n}E_nf_n \\
&+T\sum_{n}[f_n \ln f_n+(1-f_n) \ln (1-f_n)]
\end{align} 
Evidently, the free energy can be calculated if the solution of the BdG Hamiltonian is known.


\end{document}